\documentclass[prl, twocolumn, noeprint]{revtex4-1}
\pdfoutput=1

\usepackage{cmap}

\usepackage[english]{babel}
\usepackage{amsmath}

\usepackage{graphicx}
\usepackage[caption=false]{subfig}

\usepackage{hyperref}
\hypersetup{
    colorlinks=true,
    urlcolor=blue
}

\begin{document}

\title{Ultrafast reversal of the ferroelectric polarization by a midinfrared pulse}

\author{Veniamin A. Abalmasov}
\email{abalmasov@iae.nsc.ru}

\affiliation{Institute of Automation and Electrometry SB RAS, 630090 Novosibirsk, Russia}

\date{\today}

\begin{abstract}

We calculate the ferroelectric polarization dynamics induced by a femtosecond midinfrared pulse as measured in the recent experiment by R. Mankowsky et al., \href{https://journals.aps.org/prl/abstract/10.1103/PhysRevLett.118.197601}{Phys. Rev. Lett. 118, 197601 (2017)}. It is due to the nonlinear coupling of the excited infrared-active phonon with the ferroelectric mode or to the excitation of the ferroelectric mode itself depending on the pulse frequency. To begin with, we write the LiNbO$_3$ crystal symmetry invariant thermodynamic potential including electric field and nonlinear phonon coupling terms. We solve the equations of motion determined by this potential for phonon coordinates numerically in classical approximation. We explain the transient polarization reversal observed in the experiment by action of the depolarizing electric field which is due to bound charges at the polarization domain boundaries and give a reasonable estimate for its value. We argue that the polarization could be ultimately reversed when this field is screened.

\end{abstract}

\maketitle


\section{Introduction}

Control over polarization is essential for many applications of ferroelectrics, from non-volatile memory storing \cite{scott2007} to the switchable surface chemistry and catalysis \cite{kakekhani2016}. The usual way of the polarization reversal by static or pulsed electric fields is limited in speed by hundreds of picoseconds \cite{li2004}. Several proposals have been made how to switch the polarization on a time scale of picoseconds by directly exciting the ferroelectric mode with ultra-short radiation pulses \cite{fahy1994, qi2009, herchig2014}. However, they still have not been realized in practice. Though, a 90 degrees polarization rotation in a part of domains of a multi-domain ferroelectric thin film of (Ba$_{0.8}$Sr$_{0.2}$)TiO$_3$ induced by a strong single-cycle terahertz pulse was claimed recently in \cite{grishunin2017}. At the same time, ultra-short (less than 20 ps) all-optical magnetic polarization control with low heat load in transparent ferromagnetic films has already been reported \cite{stupakiewicz2017}.

Recently, it has been proposed to switch the ferroelectric polarization by resonantly exciting infrared-active phonon mode nonlinearly coupled to the ferroelectric mode \cite{subedi2015}. This approach, developed in last decades with an appearance of very high intensity lasers and called nonlinear phononics, has already proved to be successful in ultrafast lattice control \cite{foerst2011}. The follow-up experiment \cite{mankowsky2017} has indeed demonstrated a transient switching of the polarization in LiNbO$_3$ (LNO) crystal for laser pulse fluences larger than about 60 mJ/cm$^{2}$ (with the laser pulse duration of about 150 fs). One of possible explanations of the observed rapid polarization return to its initial state was the formation of uncompensated electric charges after polarization reversal in the irradiated part of the crystal \cite{mankowsky2017} which has not been taken into account theoretically \cite{fahy1994, qi2009, herchig2014, subedi2015, mankowsky2017}.

In this paper we calculate the phonon modes dynamics in conditions of the experiment \cite{mankowsky2017}. We first argue that the equation of motion for the polarization is governed by the thermodynamic potential rather than a potential obtained from ab-initio calculations for the unrelaxed crystal which was used in \cite{subedi2015, mankowsky2017}. This implies that nonlinear coupling terms must be invariant under symmetry transformations of the crystal parent group. We find biquadratic coupling constant values from the infrared phonons frequency shift at the ferroelectric phase transition which is known for LNO from ab-initio calculations. We solve numerically equations of motion for the coupled phonon modes and determine the depolarizing electric field value that corresponds better to the polarization dynamics observed in \cite{mankowsky2017}. We propose to screen this field in experiment by a metallic wire deposited on the crystal surface around the irradiated spot in order to get an ultimate polarization reversal. We also show that an ultimate reversal of polarization in conditions of the experiment \cite{mankowsky2017} is possible in our model when the ferroelectric mode is excited resonantly, though it demands very high pump fluences.

\section{Theoretical approach}
\label{sec:theor}

In crystals, the movement of interacting with each other atoms near their equilibrium positions is a superposition of a complete set of normal modes $\{Q\}$ which are plain waves with definite frequencies and polarizations and which transform according to irreducible representations of the crystal symmetry group. According to the phenomenological Landau theory, the second order phase transition takes place when the coefficient of the quadratic term of one of the coordinates in the thermodynamic potential $F(\{Q\}, T)$ becomes negative below the critical temperature $T_c$. This leads to a non-zero thermal equilibrium value of this coordinate which is proportional to the spontaneous polarization in ferroelectrics  \cite{landau2013}.

It is believed that the thermodynamic potential determines the dynamics of the order parameter as well. In the case of a displacive structural phase transition the corresponding equation is \cite{ginzburg1980}:
\begin{align}\label{motion}
   \ddot{Q} + \gamma \dot{Q} + \partial F(\{Q\}, T)/\partial Q = 0,
\end{align}
where $\gamma$ is the damping constant. In the static case this equation reduces to the usual thermal equilibrium condition. In the low-frequency range $F(\{Q\}, T)$ is calculated for fixed generalized forces conjugate to all other generalized coordinates. Thus understood Eq. (\ref{motion}), if valid up to optical phonon frequencies, expresses the essence of the so-called soft mode concept. Indeed, the square frequency in Eq. (\ref{motion}) is equal to the inverse static susceptibility $\chi^{-1}(T) = \partial^2 F(\{Q\}, T)/\partial Q^2$ which becomes zero at $T = T_c$. Only for high-frequency coordinates $F(\{Q\}, T)$ should be calculated for fixed values of the slowly changing coordinates \cite{ginzburg1980}.

We note that in \cite{subedi2015} the potential $V(\{Q\})$ which determines equations of motion similar to Eq. (\ref{motion}) was calculated for PbTiO$_3$ crystal from first principles using density functional theory for fixed values (corresponding to the low-temperature ferroelectric phase) of all the rest of coordinates. For this reason the potential $V(\{Q\})$ was not symmetric in the ferroelectric mode coordinate $Q_P$ (though the symmetry was restored when other coordinates were relaxed for a given value of $Q_P$) \cite{subedi2015}. The thus obtained potential would be appropriate for a very fast dynamics of the ferroelectric mode. At the same time, in the experiment \cite{mankowsky2017} the characteristic time change of $Q_P$ is not smaller than its inverse frequency. Moreover, the signal of the second harmonics  vanishes at some point. This implies that the crystal becomes centrosymmetric at this moment which is not possible when only the polarization vanishes but other coordinates are not relaxed with their values corresponding to the noncentrosymmetric ferroelectric phase.

\section{LNO thermodynamic potential}

\begin{figure}[t]
\center
\includegraphics[width=0.9\columnwidth]{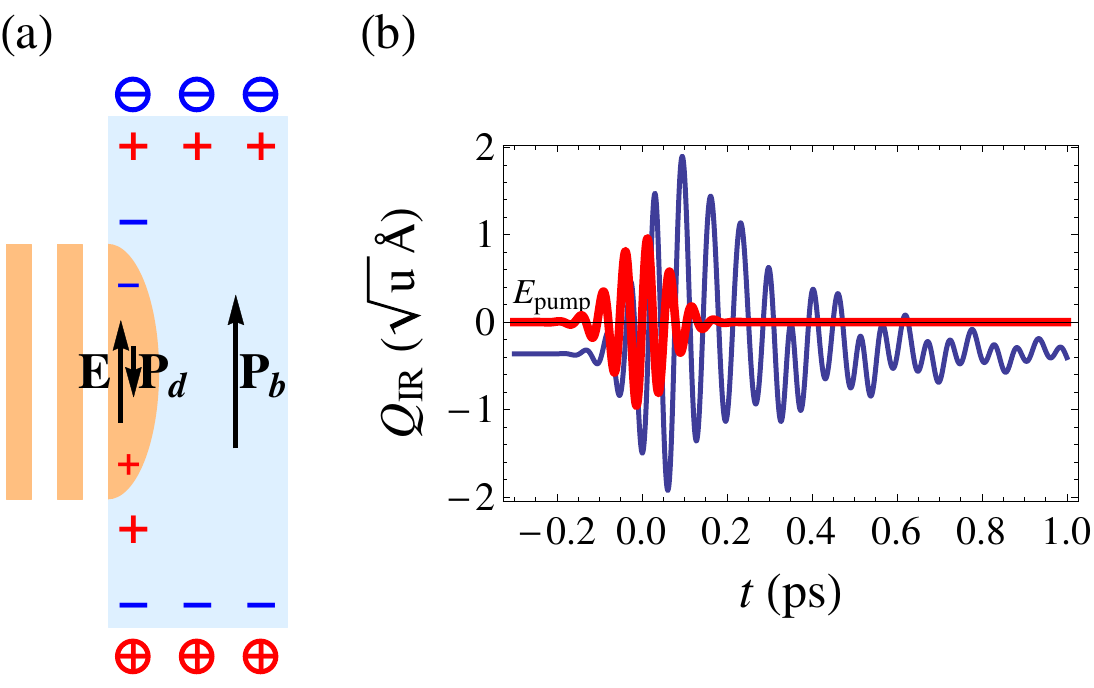}
\caption{Sketch of the bulk and domain polarization ${\bf P}_{b,d}$ and electric field $\bf{E}$ due to the bound charges in the sample when irradiated by a terahertz pulse; free charges on the surface are in circles (a) and the time-dependence of the terahertz radiation electric field (thick, red) and the corresponding infrared mode amplitude $Q_{A_1\text{(TO)}_4}$ (blue) nonlinearly coupled to the ferroelectric mode at a fluence of 95 mJ/cm$^{2}$ (b).}
\label{fig:scheme}
\end{figure}

In what follows we will consider for concreteness the LNO crystal and the experiment scheme and conditions as in \cite{mankowsky2017}, see Fig. \ref{fig:scheme}. Ferroelectric phase transition in LNO crystal occurs at the temperature about 1480 K from the paraelectric phase with symmetry $R\overline{3}c$ ($D^6_{3d}$) to the low-temperature ferroelectric phase with symmetry $R3c$ ($C^6_{3v}$) \cite{parlinski2000}. Two formula units in the unit cell implies 27 optical phonon modes, $4A_1 + 5A_2 + 9E$ in the ferroelectric phase and $A_{1g} + 2A_{1u} + 3A_{2g} + 3A_{2u} + 4E_g + 5E_u$ in the paraelectric phase. In the ferroelectric phase infrared-active polar optical phonon modes $A_1$(TO$_{1-4}$) have frequencies about 7.5, 8.1, 10 and 19 THz and $E$(TO$_{1-9}$) modes are with frequencies about 4.6, 7.0, 7.9, 9.7, 10.8, 11.1, 13.0, 17.3 and 19.8 THz \cite{kojima2016, margueron2012}. 
In the paraelectric phase $A_1$(TO$_3$) has irreducible representation $A_{1g}$ while the others are $A_{2u}$ \cite{parlinski2000}. The mode $A_1$(TO$_1$) becomes softer approaching the phase transition \cite{ridah1997} and according to the first-principles calculations \cite{veithen2002} it has the strongest overlap (0.82) with the mode $A_{2u}$ which is unstable in the paraelectric phase and coincides with the atomic displacements during the phase transition. The mixing of $A_1$(TO$_1$) and $A_1$(TO$_2$) modes at temperatures between 400 and 600 K \cite{ridah1997} is probably the cause of the incomplete overlap in the first-principles calculations. We will label the ferroelectric soft mode as $Q_P$ and the other infrared-active modes $Q_{\text{IR}}$ as $Q_{A_{1g}}$, $Q_{A_{2u}}$, $(Q_{E_{g}, x}, Q_{E_{g}, y})$ and $(Q_{E_{u}, x}, Q_{E_{u}, y})$ according to their irreducible representation.

During the ferroelectric phase transition the condensation of the soft mode occurs at the center of the Brillouin zone. The teraherz pulse as in the experiment \cite{mankowsky2017} also excites mostly long-wavelength phonons. So we will consider an homogeneous case and neglect the interaction of phonons with different wavelengths coming from nonlinear phonon coupling.

The thermodynamic potential density $F(\{Q\}, T)$ that we want to write should be invariant under transformations of the parent paraelectric phase crystal symmetry group. We will  write it as a sum
\begin{align}\label{F}
    F = F_0 + F_{E\text{-ph}} + F_{\text{ph-ph}}.
\end{align}
The first part describes free phonons:
\begin{align}\label{F0}
    F_0 =  - \frac{\omega^2_P}{4} Q_P^2 + \frac{c_P}{4} Q_P^4 + \frac{\omega_{\text{IR}}^2}{2} Q_{\text{IR}}^2,
\end{align}
where $\omega_{P, {\text{IR}}}$ are frequencies of the corresponding modes $Q_{P, {\text{IR}}}$. The coefficient $c_P$ determines the equilibrium value $Q_P^e$ through the equation $\partial F/\partial Q_P = 0$. In the absence of electric field and nonlinear phonon interactions this gives $c_P = \omega_P^2/2 (Q_P^e)^2$. Summation over all infrared-active modes $Q_{\text{IR}}$ in Eq. (\ref{F0}) is assumed.

As a function of the ferroelectric mode amplitude $F_0(Q_P)$ was calculated ab-initio, for instance, in \cite{inbar1995} and when fitted to a forth order polynomial it provided the energy difference between the ground and the lowest excited state in agreement with experimental data.

The second part of the thermodynamic potential corresponds to the phonon-electric field coupling:
\begin{align}\label{F_E-ph}
    F_{E\text{-ph}} =   & - E_z (Z^*_P Q_P + Z^*_{A_{2u}} Q_{A_{2u}} + Z_{A_{1g}} Q_P  Q_{A_{1g}}) \nonumber \\
& - \sum_{j=x,y} E_j (Z^*_{E_{u}} Q_{E_{u}, j} + Z_{E_{g}} Q_P  Q_{E_{g}, j}),
\end{align}
where $Z^*_{P, {\text{IR}}}$ are Born effective charges of the corresponding modes $Q_{P, {\text{IR}}}$, $Z_{A_{1g}, E_{g}}$ are coupling constants, $E_{x, y, z}$ are the electric field components.

We see from Eq. (\ref{F_E-ph}) that coupling of the electric field to symmetric phonon modes $A_{1g}$ and $E_{g}$ is possible only in ferroelectric phase, when $\langle Q_P \rangle \neq 0$. Though, even in ferroelectric phase this coupling is expected to be not large due to the small value of $\langle Q_P \rangle$. Indeed, the effective charge of $A_1$(TO$_3$), which has irreducible representation $A_{1g}$ in paraelectric phase, is very small \cite{kojima2016}.

We will focus on a single phonon mode excitation and as a consequence only on two-phonon modes coupling, one of each is the ferroelectric soft mode $Q_P$ and the other is an infrared-active mode $Q_{\text{IR}}$ (see \cite{juraschek2017, radaelli2018} for discussions on three phonon modes interaction). The leading coupling terms in phonon amplitudes (up to the forth order) are
\begin{align}\label{F_ph-ph}
    & F_{\text{ph-ph}}  =  \sum_{i=1,2} a_{i} Q_P^2  Q_{A_{1g}}^i + \sum_{i=1,2,3} c_{i} Q_P^{i} Q_{A_{2u}}^{4-i}  \nonumber \\
&  + \sum_{\substack{i=g,u \\ j=x,y}} b_{i} Q_P^2  Q_{E_{i}, j}^2 + d\, Q_P  Q_{E_{u}, y} (3 Q_{E_{u}, x}^2 - Q_{E_{u}, y}^2).
\end{align}

Three-phonons interaction in Eq. (\ref{F_ph-ph}) with the coupling constant $a_1$ is not of much interest to us because it is difficult to excite $A_{1g}$ mode as we discussed above (since its effective charge is proportional to $Q_P$, this term effectively is of the forth order in phonon coordinates).

In what follows we apply the thermodynamic potential (\ref{F}) through Eq. (\ref{motion}) to describe the results of the experiment \cite{mankowsky2017} where the mode $A_1$(TO$_4$) was pumped resonantly by a high-intensity femtosecond electromagnetic pulse. For this purpose we precise in the next section the numerical values of parameters which enter Eqs.~(\ref{motion})~-~(\ref{F}).

\section{Values of parameters}

The electric field in Eq. (\ref{F_E-ph}) has two constituents, $E_z = E_1 + E_2$, the both being directed along $z$-axis in the experiment \cite{mankowsky2017}. One is the driving midinfrared pulse electric field $E_1 (t) = E_0 \sin(\omega t) \exp(-4\ln2 \,t^2 / T^{2})$ with frequency $\omega$, Gaussian envelope of duration $T = 0.15$~ps and amplitude $E_0$ up to 25 MV/cm in the experiment \cite{mankowsky2017}, see Fig. \ref{fig:scheme}(b). The other component is due to the depolarizing electric field $E_d$ in the reversed polarization domain created by the terahertz pump (Fig.~\ref{fig:scheme}(a)). We suppose there is no screening of this field by free carriers on the time scale of the polarization reversal as in the experiment \cite{mankowsky2017}. So the resulting field is $E_2(t) = E_d (1 - Q_P(t)/Q_P^e)$.

We calculate the effective electric charge of a given optical mode from the experimental value of its oscillation strength \cite{hoegen2018}. Thus, we obtain $Z^*_P= 1.356$, $Z^*_{A_1(\text{TO}_2)} = 0.564$ and $Z^*_{A_1(\text{TO}_4)} = 1.404$ $e/\sqrt{\text{u}}$.  The effective charge $Z^*_{A_1(\text{TO}_3)}$ is very small \cite{ridah1997} and we will neglect the dynamics of this mode.

The ions shifts between paraelectric and ferroelectric phase at room temperature according to \cite{lines2001} correspond to $Q_P^e = 2.9 \sqrt{\text{u}}${\AA} which agrees with the minimum position of the two-minimum energy surface calculated in \cite{inbar1995}. At the same time, according to the experimental data \cite{boysen1994} the Li and O atoms are shifted in the ferroelectric phase by about $\Delta z_{\text{Li}} = 0.460$ {\AA} and $\Delta z_{\text{O}} = 0.270$~{\AA} which gives the amplitude $Q_P^e = (\sum_i m_i \Delta z_i^2)^{1/2} = 3.1 \sqrt{\text{u}}${\AA}. The same value of $Q_P^e$, which we adopt in our calculations, follows from atomic displacements calculated ab-initio \cite{parlinski2000, veithen2002, friedrich2015}. $Q_P^e$ enters Eq. (\ref{motion}) as an initial value of the $Q_P(t)$ coordinate. Initial values of other coordinates are obtained from the equilibrium condition $\partial F / \partial Q_{\text{IR}} = 0$ and they are not zero when the modes are nonlinearly coupled, see Fig. \ref{fig:scheme}(b). This also can explain a partial overlap of ferroelectric soft modes in para- and ferroelectric phases as it was discussed above.

We note that the spontaneous polarization $P_s = Z_P^*Q_P^e/v_0$, $v_0$ being the unit cell volume, calculated for $Q_P^e = 3.1 \sqrt{\text{u}}${\AA} is about $0.64$ C/m$^{2}$ and slightly lower than the experimental value about $P_s = 0.70$ C/m$^{2}$ \cite{veithen2002} which in its turn is attained for a rather large value $Q_P^e = 3.4 \sqrt{\text{u}}${\AA}. This slight discrepancy can also be attributed to the highly nonlinear evolution of the charges along the ferroelectric path of atomic displacements \cite{veithen2002}.

We adopt the values for the damping constants in Eq. (\ref{motion}) from \cite{hoegen2018} to be $\widetilde{\gamma}_P= 0.8$, $\widetilde{\gamma}_{A_1(\text{TO}_2)} = 0.6$ and $\widetilde{\gamma}_{A_1(\text{TO}_4)} = 1.0$ THz (with $\gamma = 2 \pi \widetilde{\gamma}$). The damping $\widetilde{\gamma}_P$ increases strongly with temperature and equals the soft mode frequency of 5 THz at about 1100 K \cite{ridah1997}. It is not clear whether this is due to the mode softening or just to the temperature dependence. We keep the damping $\widetilde{\gamma}_P$ constant in our calculations.

Finally, we note that the biquadratic phonon-phonon interaction with the coupling constant $c_2$ in Eq. (\ref{F_ph-ph}) does change not only the frequency of the soft mode $Q_P$ but the frequency of the $Q_{\text{IR}}$ as well, Eq. (\ref{F0}). This allows, for instance, in some cases to reproduce the polarization temperature dependence from high-frequency modes temperature dependence \cite{krylov2013, salje1997}. Usually the effect is not large. For the ferroelectric KDP crystal, however, the change is about ten percent for certain modes indicating both signs of the coupling constant \cite{brehat1987, simon1988}. So we calculate the coupling constant as $c_{2} = (\omega_{\text{IR}}^2 - \Omega_{\text{IR}}^2)/2  (Q_P^e)^2$ where $\omega_{\text{IR}}$ and $\Omega_{\text{IR}}$ are the frequencies of the $Q_{\text{IR}}$ mode in the ferroelectric and paraelectric phases respectively, $Q_P^e$ is the thermal equilibrium value of $Q_P$ at room temperature. The positive sign of the coupling constant assures a single well potential $F(Q_P)$ for large values of $Q_{\text{IR}}^2$ and thus a possibility of the polarization reversal. We note that $\omega_{\text{IR}}$ does not change substantially in the ferroelectric phase of LNO up to about 1000 K \cite{johnston1968, ridah1997} but this agrees with a small change in the polarization itself in this temperature range \cite{shostak2009}.

Due to the very high temperature of the phase transition the values of $\Omega_{\text{IR}}$ for LNO crystal are available only from first-principles calculations. For $A_1$(TO$_4$) they vary from 15.6 \cite{parlinski2000} and 14.3 \cite{veithen2002} to 13.6 THz \cite{friedrich2015},  for $A_1$(TO$_2$) from 3.5 \cite{parlinski2000} and 2.8 \cite{veithen2002} down to 0.9 THz \cite{friedrich2015}. For $A_1$(TO$_3$) in contrast the frequency is larger in the paraelectric phase varying from 12.1 \cite{parlinski2000, veithen2002} to 10 THz \cite{friedrich2015} which indicates a possible negative value of the biquadratic phonon-phonon coupling constant. We adopt values $\Omega_{A_1(\text{TO}_4)} = 14$ THz and $\Omega_{A_1(\text{TO}_2)} = 3.5$~THz which correspond to $c_{2, A_1(\text{TO}_4)} = 34$ meV/u$^{2}${\AA}$^{4}$ and $c_{2, A_1(\text{TO}_2)} = 11$ meV/u$^{2}${\AA}$^{4}$. We note that our coupling constants appear to be an order of magnitude smaller than those obtained for quantum paraelectric crystals KTaO$_3$ \cite{subedi2017} and  SrTiO$_3$ \cite{kozina2019} from DFT calculations. This might be due to the calculation procedure of the potential discussed in Sec. \ref{sec:theor}. Indeed, a relaxed lattice has the lowest energy and, as a consequence, the nonlinear phonon coupling constants of our thermodynamic potential are smaller.

We keep coupling constants $c_1$ and $c_3$ in Eq. (\ref{F_ph-ph}) and the depolarizing electric field $E_d$ in Eq. (\ref{F_E-ph}) as fitting parameters in our calculations when compared to the experimental results \cite{mankowsky2017}.

\begin{figure}[t]
\center
\includegraphics[width=0.8\columnwidth]{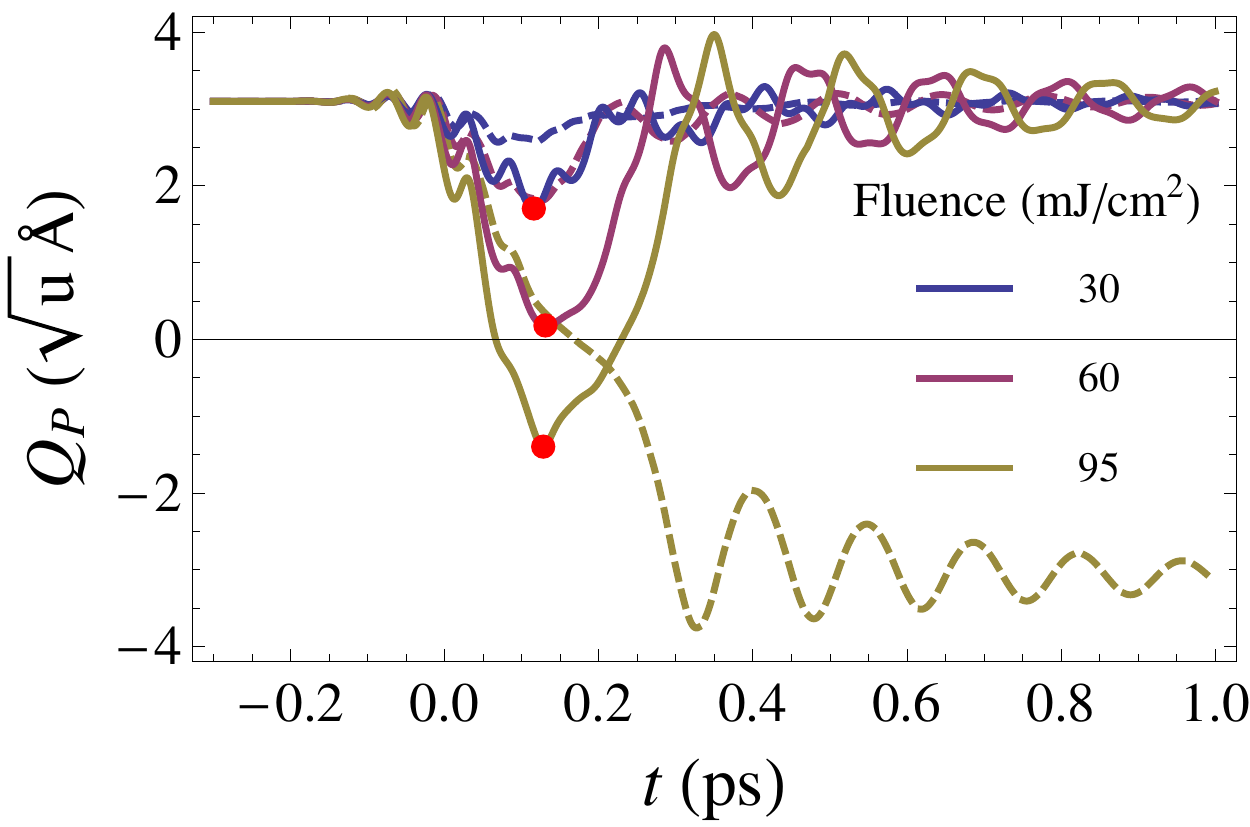}
\caption{Ferroelectric mode dynamics for the pump pulse fluences of 30, 60 and 95 mJ/cm$^{2}$ and frequency $\omega =  19$ THz when $c_{1, 3} = 0$ and $E_d = 0$ (dashed) and $c_1 = 10$, $c_3 = 17$ meV/u$^{2}${\AA}$^{4}$ for $A_1(\text{TO}_4)$ and $E_d = 2.8$ MV/cm (solid). Red dots are at minimums of $Q_P$ }
\label{fig:QP}
\end{figure}

\section{Calculation results}

We start with zero values of coupling constants $c_{1, 3}$ and the depolarizing electric field $E_d$ and calculate the phonon modes dynamics for three values of the infrared pulse fluence, see Fig.~\ref{fig:QP}. In the experiment \cite{mankowsky2017} for fluences above a threshold value of 60 mJ/cm$^{2}$ the second-harmonic intensity (and thus the ferroelectric mode $Q_P$) was observed to vanish completely. We see, however, only a reduction of it at this fluence. The dynamics becomes closer to the experiment if we put $c_{2, A_1(\text{TO}_4)} = 51$~meV/u$^{2}${\AA}$^{4}$ but this corresponds to the very low frequency $\Omega_{A_1(\text{TO}_4)} = 10.7$ THz expected in paraelectric phase. Interestingly, the situation can be improved if we take a positive value of the constant $c_{3, A_1(\text{TO}_4)} \gtrsim 13$~meV/u$^{2}${\AA}$^{4}$. It is rather unexpected because the force exerted by this coupling on $Q_P$ oscillates and changes its sign. At the same time, negative values of $c_{3, A_1(\text{TO}_4)}$ make the polarization reversal even harder. Finite values of $c_{1, A_1(\text{TO}_4)}$ do not influence much the dynamics. This can be easily understood because the latter coupling is cubic in $Q_{\text{IR}}$ while the former is linear and $Q_{\text{IR}} \ll Q_P$.

The reentrant behavior of the polarization for the largest fluence of 95 mJ/cm$^{2}$ available in \cite{mankowsky2017} appears for depolarizing electric fields larger than about $E_d \approx 2.5$~MV/cm. This value to be compared with the depolarizing field in a plate-like monodomain sample $E_{d} = P_s / (\varepsilon\varepsilon_0)$ which is about 26.4 MV/cm for $P_s= 0.70$ C/m$^{2}$ and the dielectric constant $\varepsilon_{33} = 30$ in LNO ($\varepsilon_0$ is the electric constant). The value of the depolarization factor $N \approx 0.1$ seems to be reasonable taken into account the oblong shape of the reversed polarization domain created (the pump penetration depth is about 3.2 $\mu$m \cite{mankowsky2017} and its spot size is about 65 $\mu$m \cite{hoegen2018}, see Fig. \ref{fig:scheme}(a)).

Finally, we adopt the values $E_d = 2.8$~MV/cm and $c_1 = 10$ and $c_3 = 17$ meV/u$^{2}${\AA}$^{4}$ for $A_1(\text{TO}_4)$ mode, see Fig.~\ref{fig:QP}. For $A_1(\text{TO}_2)$ we keep these coupling constants zero since its dynamics (due to the small Born electric charge $Z^*_{A_1(\text{TO}_2)}$) does not influence visibly the dynamics of $Q_P$ even near the resonance which is close to the resonance of $Q_P$ itself.

\begin{figure}[t]
\centering
\includegraphics[width=0.8\columnwidth]{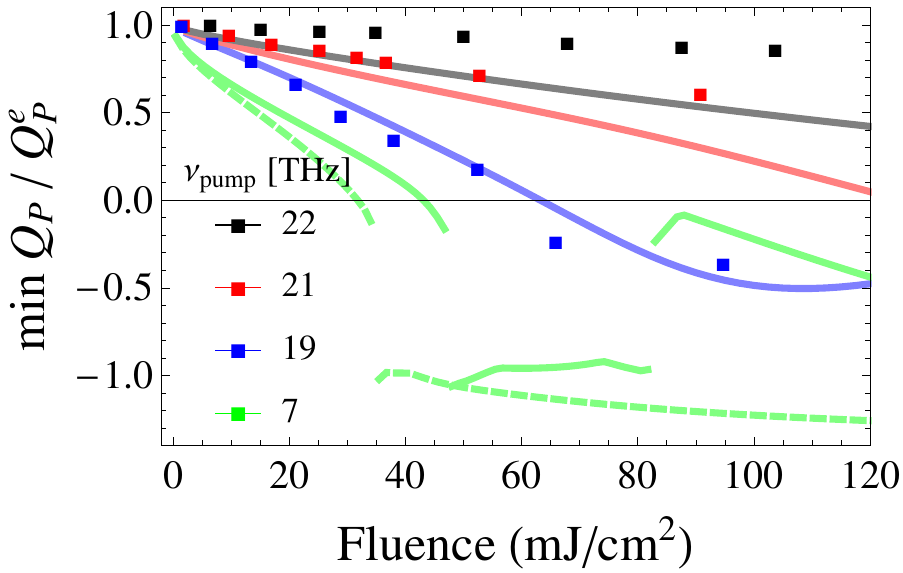}
\caption{Minimum of $Q_P$ (normalized by $Q_P^e$) as a function of fluence for different frequencies of the pump pulse. At $\nu_{\text{pump}} = 7$ THz two lines correspond to opposite signs of the pump pulse electric field. Experimental data are from \protect\cite{mankowsky2017}}
\label{fig:minQP}
\end{figure}

\begin{figure}[]
\centering
\includegraphics[width=0.8\columnwidth]{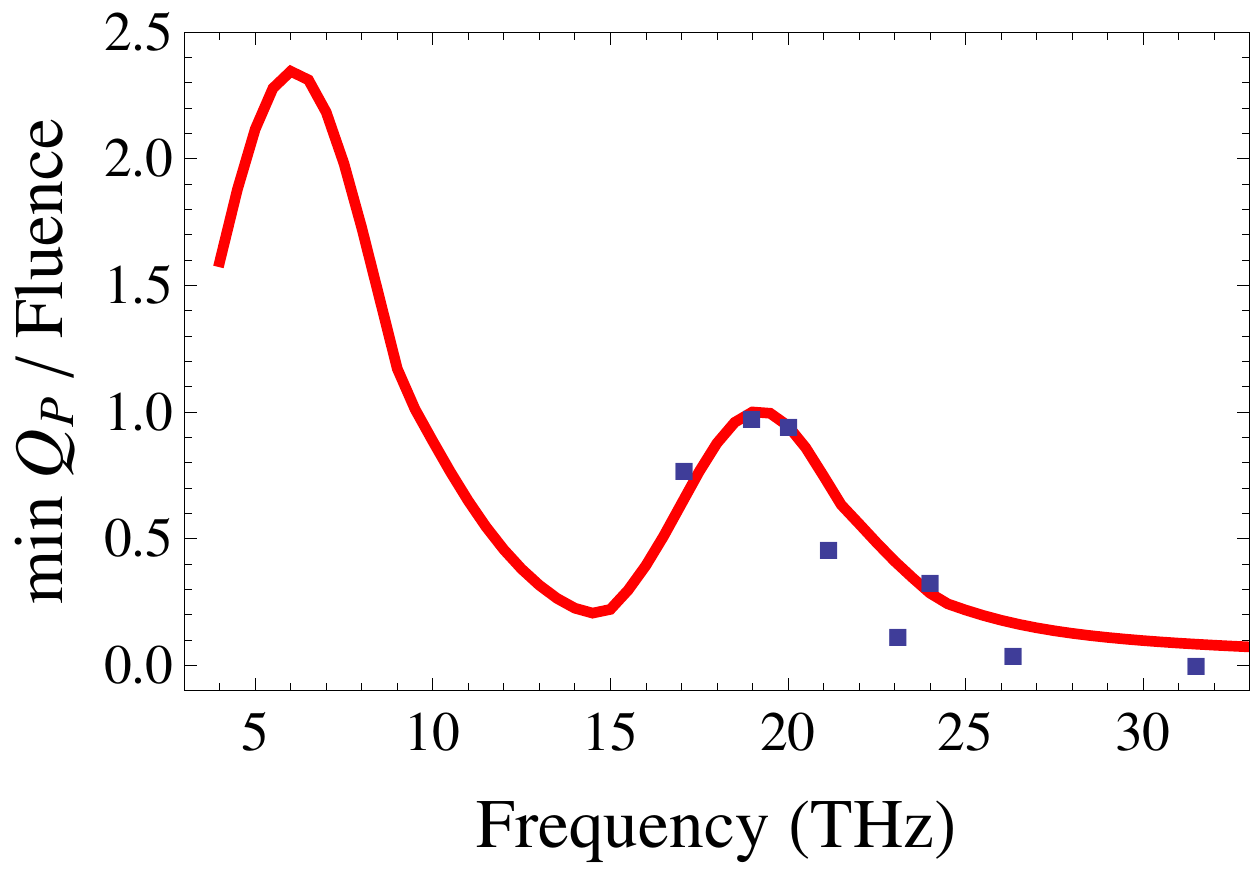}
\caption{Susceptibility as a function of the pump frequency normalized to unity at 19 THz. Experimental data are from \protect\cite{mankowsky2017}}
\label{fig:suscept}
\end{figure}

We calculate the minimum value of $Q_P$ (normalized by $Q_P^e$) as a function of fluence for different frequencies of the pump pulse (with the same Gaussian envelope duration) and compare our results to the experiment \cite{mankowsky2017}, see Fig. \ref{fig:minQP}. The agreement is good enough. For the pump frequency of 7 THz, which is close to the ferroelectric mode resonance, the result depends slightly on the sign (phase) of the pulse oscillations due to comparable values of the pump duration and the period of oscillations. For this pump frequency we see an ultimate reversal of the polarization for fluences higher than about 40 mJ/cm$^{2}$. The reversal becomes even harder for larger values of the depolarizing field until it becomes impossible if $E_d \gtrsim 5$~MV/cm. This threshold corresponds to the coercive field value in our model, $E_{c} = P_s / (3\sqrt{3} \varepsilon\varepsilon_0)$ in the absence of electric field and nonlinear phonons coupling, at which the metastable state (local minimum) with opposite polarization disappears. We note, however, that it is an order of magnitude larger than experimental values of the coercive field in this crystal \cite{volk2008}.

The dependence of the susceptibility $\text{min} Q_P / \text{fluence}$ calculated for small fluences as a function of the pump frequency reproduces the experimental results \cite{mankowsky2017} as well (Fig. \ref{fig:suscept}). The width of peaks is determined mostly by the frequency width of the pump pulse.

\section{Discussion}

In our calculations we see the oscillations of $Q_P$ with the ferroelectric mode frequency which are absent in the experiment \cite{mankowsky2017}. In the experiment these oscillations could be smeared for several reasons. First, initial values of velocities are not zero and those of coordinates are not at equilibrium but they are determined instead by the temperature and the coordinates wave functions. Second, the electric field amplitude of the midinfrared pulse which penetrates the crystal is not homogeneous and is determined by the Gaussian function perpendicular to its direction and a vanishing exponential deep into the crystal, $\exp(-z/z_{0})$ with $z_{0} \approx 3.5 \,\mu$m, while the second harmonic is generated at $z \lesssim 1 \,\mu$m \cite{mankowsky2017}. Finally, the nonlinear coupling to the phonon modes with nonzero wave numbers which we have not taken into account would probably also lead to the polarization oscillations smearing.

\begin{figure}[t]
\centering
\includegraphics[width=0.98\columnwidth]{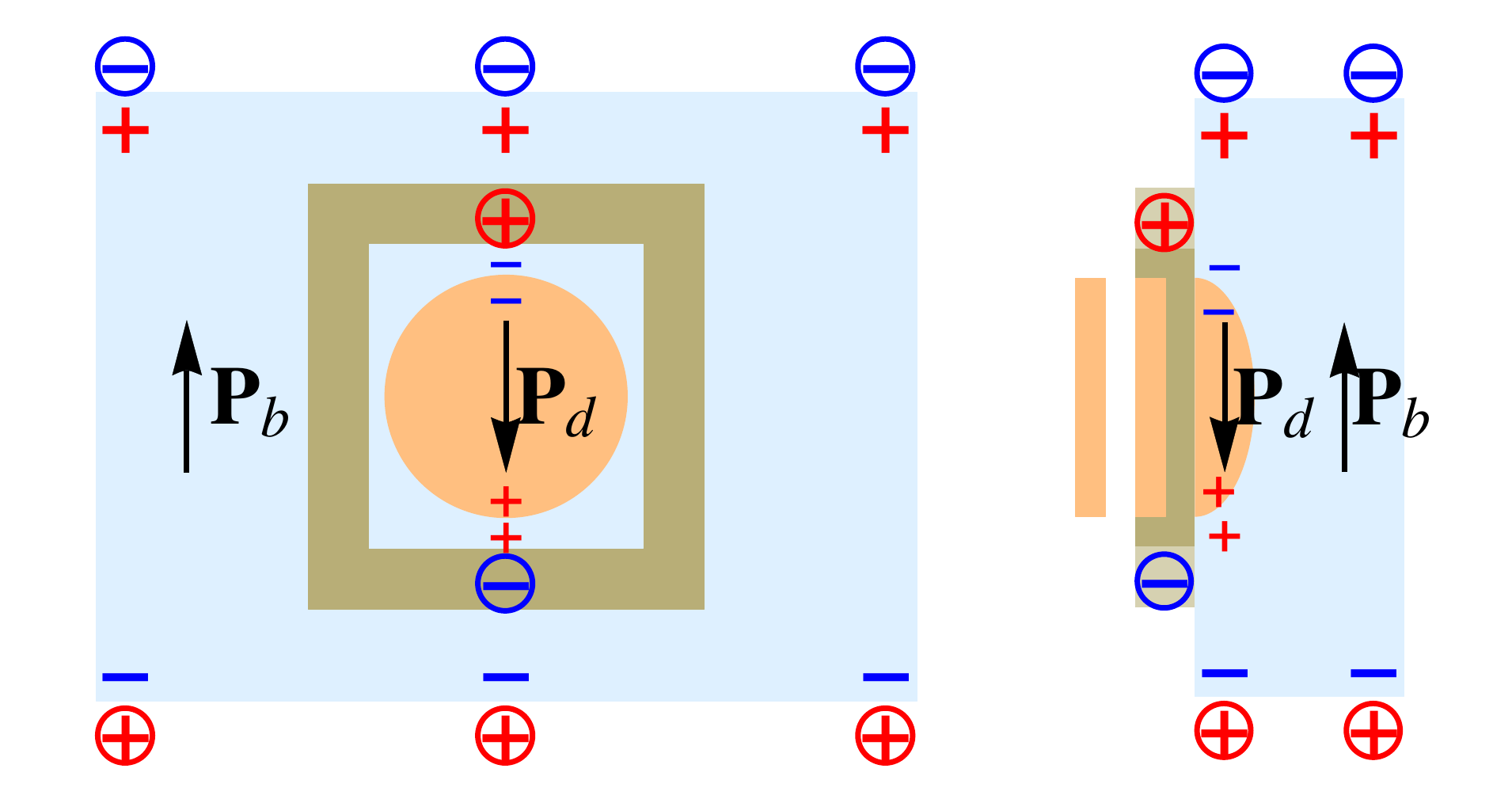}
\caption{Sketch of the screen (in golden) on the top of the crystal around the pump laser spot and the electric charges distribution (free charges are in circles)} 
\label{fig:screen}
\end{figure}

High-intensity sources of far-infrared electromagnetic fields (more than 3.5 MV/cm) tunable between 4 and 18 THz were reported recently in \cite{liu2017}. These fluences, however, are not high enough to switch the polarization by the resonant pump of the ferroelectric mode (Fig. \ref{fig:minQP}). A very high-intensity source of terahertz single-cylce pulse (up to 22 MV/cm) was reported in \cite{ovchinnikov2019} but its frequency is about 1 THz only. Thus, the resonant ferroelectric mode reversal in LNO seems to be unattainable at the moment.

Our calculations show, however, that the polarization can be ultimately switched in the absence of the depolarizing field (Fig. \ref{fig:QP}). In order to screen this field a metallic wire could be used which is drawn in golden in Fig. \ref{fig:screen}. The relaxation time of this screen $\tau = R C$ is determined by the resistance $R \approx  \rho L / (h + d) \delta$ and capacitance $C \approx \varepsilon\varepsilon_0 h$ of the wire. The resistivity and skin depth for gold at frequency 10 THz are $\rho \sim 10 ^{-7} \,\Omega$~m and $\delta \approx 30$ nm. The metallic wire length, height and width are about $L \approx 400 \,\mu$m, $h \sim d \sim 10 \div 100 \,\mu$m. This yields the relaxation time smaller than $\tau \sim 0.1$ ps, which is enough to follow the polarization dynamics (Fig. \ref{fig:QP}).

Recently, anharmonic oscillations of the lowest frequency $E$ mode in LNO were studied under resonant excitation by a single-cycle pulse with the electric field about 1 MV/cm \cite{dastrup2017}. At the same time, it would be interesting to see the dynamics of the ferroelectric mode $Q_P$ induced by the resonant excitation of the $E$ modes, Eqs. (\ref{F_E-ph})-(\ref{F_ph-ph}), due to their nonlinear coupling as it was described in this work for the $A$ modes excitation, especially keeping in mind orthogonal polarizations of the $A$ and $E$ modes. Calculated ab-initio $E$ modes frequencies \cite{parlinski2000, friedrich2015} differ substantially in para- and ferroelectric phases meaning possible strong coupling constants of these modes to the ferroelectric mode.

\section{Conclusion}

Our calculations of the ferroelectric mode dynamics in LNO, determined by the parent phase symmetry-invariant thermodynamic potential, reproduce well the transient reversal of polarization under high-frequency mode excitation reported in \cite{mankowsky2017} when the electric field of the bound charges is taken into account. The estimated strength of this field agrees with the polarization value in LNO and the expected depolarization factor of the transiently created polarization domain. We argue that the polarization could be ultimately reversed if the depolarizing field is screened, for example by the metallic wire on the top of the crystal around the pump laser spot. We preview that the same dynamics of the polarization could be probed by the resonant excitation of the $E$ modes which are orthogonal to the polarization.

\section*{Acknowledgments}

I thank Roman Mankowsky and Alaska Subedi for useful discussions. 

The study was carried out with the financial support of the Russian Foundation for Basic Research in the framework of the scientific project No. 18-02-00399.

\hspace{1cm}


%

\vspace{0.5cm}

\end{document}